\documentclass[conference]{IEEEtran}
\usepackage{graphicx}
\usepackage{amsmath}
\usepackage{booktabs}
\usepackage{multirow}
\usepackage{xcolor}

\usepackage[ruled,vlined,linesnumbered]{algorithm2e}
\SetKwInput{KwIn}{Input}
\SetKwInput{KwOut}{Output}
\SetKw{KwRet}{return}
\usepackage{tikz}
\usetikzlibrary{arrows.meta}

\usepackage{graphicx}   
\usepackage{subcaption} 
\usepackage{booktabs} 
\usepackage{amssymb}   
\usepackage{pgfplots}
\pgfplotsset{compat=1.18}
\usepackage{booktabs, makecell, xcolor}
\usepackage{pgfplots}
\usepackage{cite} 
\usepackage{svg}
\usepackage{py}
\usepackage[ruled,vlined]{algorithm2e} 
\usepackage{pgfplots}
\pgfplotsset{compat=1.18}
\usepackage{amsmath}
\usepackage[flushleft]{threeparttable}
\usepackage[font=small,labelfont=bf,tableposition=top]{caption}
\newlength{\bibitemsep}
\newlength{\bibparskip}
\let\oldthebibliography\thebibliography
\renewcommand\thebibliography[1]{%
  \oldthebibliography{#1}%
  \setlength{\parskip}{\bibitemsep}%
  \setlength{\itemsep}{\bibparskip}%
}
\usepackage{xcolor}
\usepackage{pgfplots}
\usepackage{array}
\usepackage{fancybox}
\usepackage{todonotes}
\usepackage{tikz}
\usepackage{textcomp}
\usepackage{todonotes}
\usepackage[utf8]{inputenc}

\usepackage{listings}
\usepackage{color}

\selectcolormodel{rgb}

\definecolor{cbxgreenA}    {RGB}{230, 248, 219}
\definecolor{cbxgreenB}    {RGB}{190, 227, 169}
\definecolor{cbxgreenC}    {RGB}{ 88, 171,  30}
\definecolor{cbxgreenD}    {RGB}{ 42,  76,  19}
\definecolor{cbxbluegreen} {RGB}{ 19,  76,  51}
\definecolor{cbxblueA}     {RGB}{223, 238, 255}
\definecolor{cbxblueB}     {RGB}{183, 206, 233}
\definecolor{cbxblueC}     {RGB}{ 69, 136, 214}
\definecolor{cbxblueD}     {RGB}{ 20,  50,  85}
\definecolor{cbxbrownA}    {RGB}{247, 243, 204}
\definecolor{cbxbrownB}    {RGB}{213, 205, 124}
\definecolor{cbxbrownC}    {RGB}{143, 132,  19}
\definecolor{cbxbrownD}    {RGB}{ 84,  67,   0}
\definecolor{cbxredA}      {RGB}{254, 217, 217}
\definecolor{cbxredB}      {RGB}{223, 167, 159}
\definecolor{cbxredC}      {RGB}{176,  24,  24}
\definecolor{cbxredD}      {RGB}{ 98,   9,   9}
\definecolor{cbxblueitem}  {RGB}{ 40, 100, 150}

\colorlet{cbxschemeA}{cbxblueA}
\colorlet{cbxschemeB}{cbxblueB}
\colorlet{cbxschemeC}{cbxblueC}
\colorlet{cbxschemeD}{cbxblueD}

\definecolor{dmlgreen}    {RGB}{51,  160,  44}
\definecolor{dmlblue}     {RGB}{31,  120, 180}
\definecolor{dmlred}      {RGB}{202,   0,  32}

\lstdefinestyle{simple}{%
  language=Python,
  numbers={none},
  basicstyle={\ttfamily},
  moredelim={[is][\underbar]{__}{__}}
}

\definecolor{codegreen}{rgb}{0,0.6,0}
\definecolor{codegray}{rgb}{0.5,0.5,0.5}
\definecolor{codepurple}{rgb}{0.58,0,0.82}
\definecolor{backcolour}{rgb}{0.95,0.95,0.92}

\lstdefinestyle{mystyle}{
	backgroundcolor=\color{white},
	keywordstyle=\color{codegreen},
	numberstyle=\tiny\color{codegray},
	stringstyle=\color{codepurple},
	basicstyle=\ttfamily,
	breakatwhitespace=false,
	breaklines=true,
	captionpos=b,
	keepspaces=true,
	numbers=left,
	numbersep=5pt,
	showspaces=false,
	showstringspaces=false,
	showtabs=false,
	tabsize=2
}

\usepackage{graphics}

\usepackage{float}
\usepackage{mdwlist}

\usepackage{cases}

\usepackage{multirow}

\usepackage{blindtext}

\usepackage{etoolbox}

\makeatother
\usepackage{enumitem}
\usetikzlibrary{patterns}
\usepackage{gensymb}

\usepackage{tikz}

\usepackage[
singlelinecheck=false 
]{caption}

\usepgfplotslibrary{patchplots}
\usetikzlibrary{pgfplots.statistics}
\usetikzlibrary{snakes}
\usepackage{graphics}
\usepackage{gensymb}
\usepackage{multirow}

\usepackage{xspace}
\usepackage{soul}

\usepackage{pifont}

\usepackage[symbol]{footmisc}

\usepackage{bbding}
\usepackage{makecell}
\usepackage[symbol]{footmisc}

\usepackage{array}
\usepackage{multirow}

\usepackage{xcolor}
\definecolor{myorange}{RGB}{230,126,34}
\usepackage[colorlinks=true, linkcolor=myorange, citecolor=myorange, urlcolor=myorange]{hyperref}
\def\lmtt@use@light@as@normal{}

\begin{document}



\title
{\huge
  MACO: A Multi-Agent LLM Framework for Automated CGRA Hardware/Software Co-Design \vspace{-0.4cm}
}



\author{
Zesong Jiang$^{1}$ \quad
Yuqi Sun$^{2}$ \quad
Qing Zhong$^{3}$ \quad
Mahathi Krishna$^{1}$ \\
Deepak Patil$^{1}$ \quad
Cheng Tan$^{1,3}$ \quad
Jeff Zhang$^{1}$ \\[1mm]
$^{1}$Arizona State University \quad
$^{2}$Independent \quad
$^{3}$Google \\
\texttt{\{zjian137, mravish2, dppatil3, chengtan, jeffzhang\}@asu.edu} \\
\texttt{yuqi.sun@outlook.com} \quad
\texttt{\{qingzhong, chengtan\}@google.com}
}

\maketitle

\vspace{-20pt}
\begin{abstract}
Designing optimal Coarse-Grained Reconfigurable Arrays (CGRAs) requires navigating a vast, interdependent hardware/software space bottlenecked by costly manual iteration. We present \textbf{MACO}, an open-source, multi-agent LLM framework that automates CGRA HW/SW co-design. MACO decomposes the design loop into four collaborative stages, \textit{HW/SW Co-design}, \textit{Error Correction}, \textit{Best-Design Selection}, and \textit{Evaluation\&Feedback}, to iteratively optimize power, performance, and area (PPA). To accelerate convergence and efficiently traverse the design space, MACO introduces an exponentially decaying exploration strategy, EDA-guided LLM self-learning, and robust rule-based error correction. Evaluated against state-of-the-art baselines, MACO reduces power consumption by 25.9\%, improves performance by 20.0\%, and accelerates the search process by 5$\times$. Finally, we validate MACO's physical design through a complete 7nm ASIC design flow.
\end{abstract}


\section{Introduction}\label{sec-intro}
CGRAs are a versatile computing paradigm bridging the fixed efficiency of domain-specific accelerators and the flexibility of general-purpose processors. By enabling functional-unit level reconfiguration, CGRAs adapt its hardware to application needs. This makes them highly effective for compute- and communicate-intensive workloads, 
driving their adoption across machine learning~\cite{qin2025picachu,abarajithan2026cgra4ml,li2026transmap,tan2023vecpac}, high-performance computing~\cite{jokai2024fused,wijerathne2019cascade,koeplinger2018spatial,amoeba}, and embedded systems~\cite{tan2024iced,prabhakar2017plasticine,chin2017cgra}.

\begin{figure}[t]
    \centering
    \includegraphics[width=1.0\linewidth]{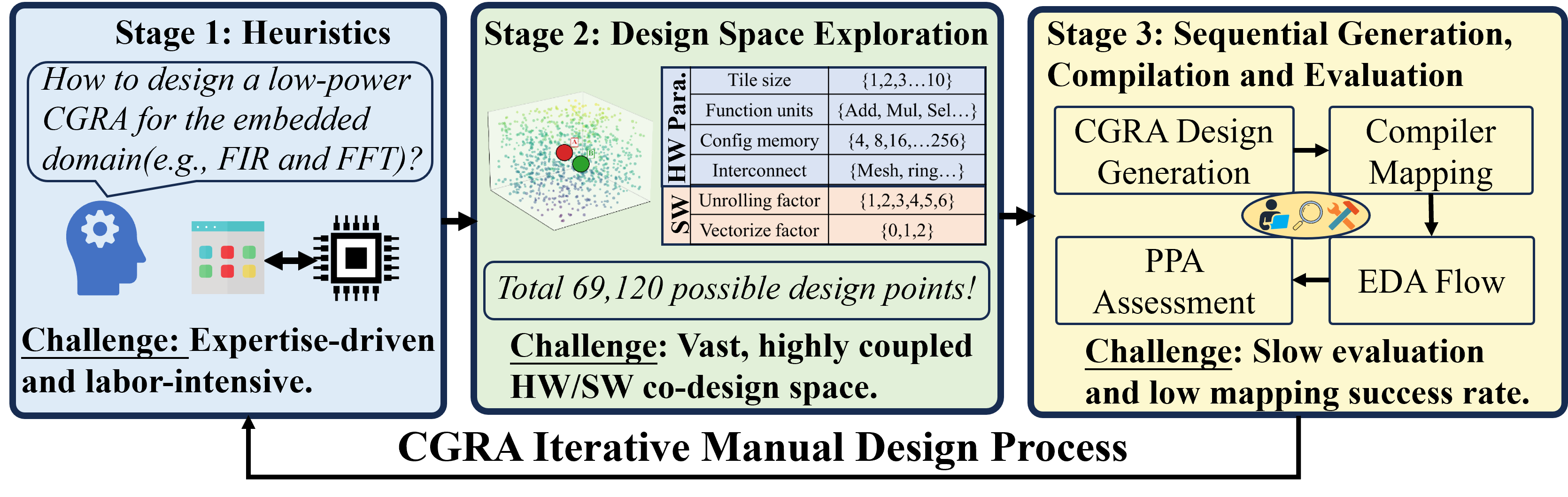}
    \vspace{-15pt}
    \caption{The state-of-the-art CGRA design flow~\cite{tan2021opencgra} and its challenges.}
    \vspace{-20pt}\label{fig:challenges}
\end{figure}

Despite these advantages,  efficient CGRA design remains challenging~\cite{tan2022asap,tan2021opencgra}. As shown in Fig.~\ref{fig:challenges}, CGRA HW/SW co-design process is bottlenecked by
three tightly coupled issues: (1) the process is highly labor-intensive and relies heavily on manual, expertise-driven reasoning; 
(2) it requires navigating a vast, interdependent joint HW/SW design space spanning parameters like tile size, functional unit and memory allocation, interconnects, loop unrolling, vectorization, and mapping strategies. Without careful design selection, it may conflict power, performance, and area (PPA) constraints;
and (3) each candidate design requires an expensive evaluation loop (design generation, compilation\&mapping, EDA flow, and PPA assessment) with notoriously low mapping success rates. Consequently, manual methodologies are unscalable and incapable of comprehensively exploring the co-design space.

\begin{figure}[t]
    \centering
    \includegraphics[width=\linewidth]{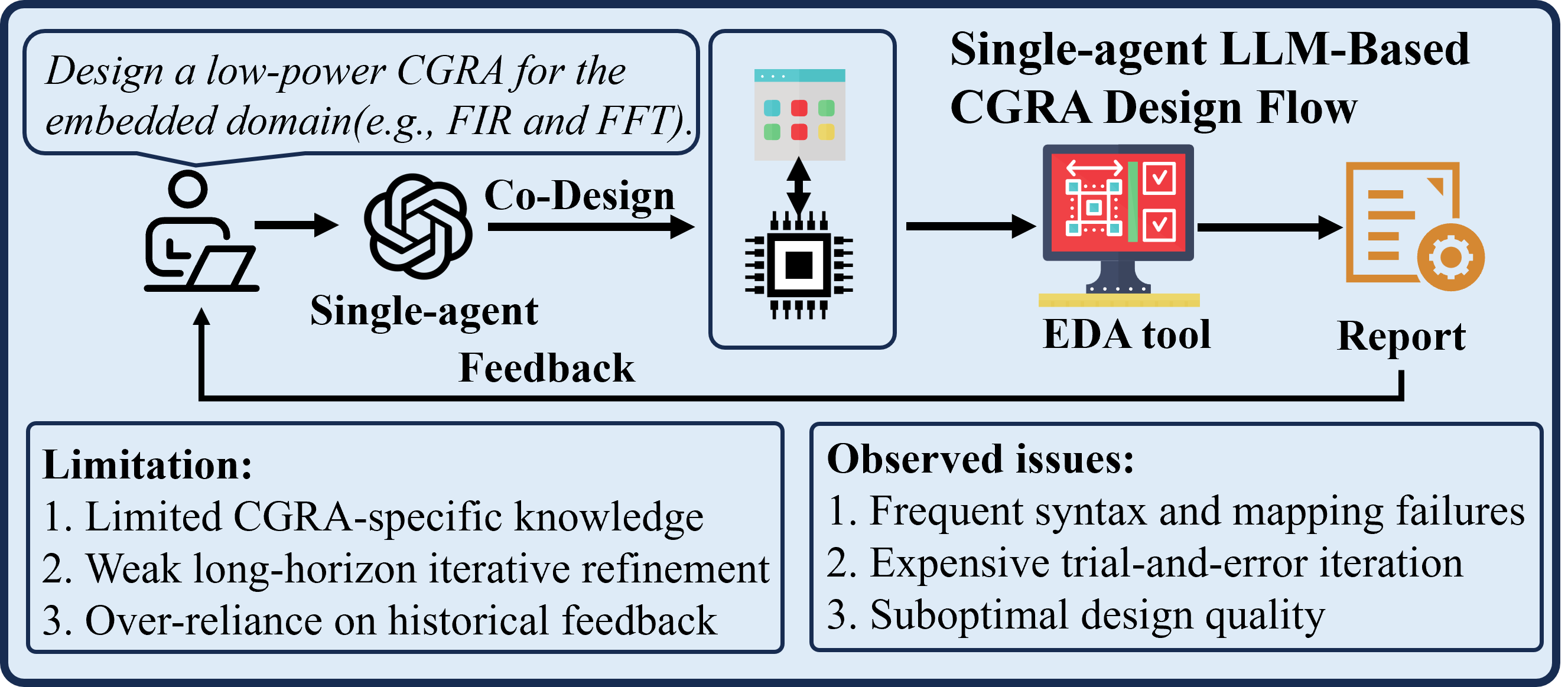}
     \vspace{-15pt}
    \caption{The single-agent LLM-based design flow\cite{du2026survey,huang2024understanding,li2026single}.}
    \vspace{-15pt}\label{fig:limitations}
\end{figure}

Recent advances in large language models (LLMs)~\cite{pan2025survey} offer a promising avenue for automated hardware design by leveraging their strong reasoning and cross-layer knowledge integration capabilities.
However, directly applying LLMs \emph{cannot} adequately address the intricacies of CGRA design.
As shown in Fig.~\ref{fig:limitations}, a typical single-agent LLM workflow relies on an iterative generate--verify--refine loop. This monolithic approach suffers from critical intrinsic limitations: (1) due to a lack of deep, domain-specific CGRA knowledge, generated designs frequently suffer from syntax and mapping failures; (2) requiring a single agent to handle a long-horizon refinement process involving generation, verification, error correction, and optimization is highly inefficient and often degenerates into expensive trial and error; and (3) as the agent relies heavily on its own historical reports, the search often biases toward past design patterns, leading to suboptimal local minima.

To overcome these limitations, we propose MACO, an open-source multi-agent LLM-based CGRA HW/SW co-design flow. Rather than relying on a monolithic loop, MACO explicitly decomposes the long-horizon co-design process into four coordinated, stage-specific agents: \textit{HW/SW co-design}, \textit{Design Error Correction}, \textit{Best Design Selection}, and \textit{Evaluation \& Feedback}. 
This decomposition localizes distinct challenges within the workflow and applies targeted mechanisms to solve them.
To sustain effective search across the vast design space, MACO employs a novel exploration-exploitation coordination strategy that mitigates self-reinforcing loops and accelerates convergence.
We further integrate a rule-based mechanism for robust error correction and a self-learning paradigm that enables the LLM to progressively acquire PPA insights from EDA tool reports,
allowing it to predict design quality and select optimal candidates.

Experimental results across diverse workloads demonstrate that MACO efficiently generates high-quality CGRA designs while drastically reducing manual design effort. Our main contributions are summarized as follows:
\begin{itemize}[noitemsep, topsep=0pt, parsep=0pt, partopsep=0pt, leftmargin=1em]
  \item We propose MACO, the first open-source multi-agent hardware/software framework for CGRA design. 
  By utilizing tightly integrated, LLM-driven agents to manage compilation, architectural exploration, simulation\&verification, MACO enables holistic and automated exploration of the vast HW/SW design space. 
  \item We introduce three novel mechanisms to enhance the agents' capabilities: an exploration-exploitation coordination strategy to avoid local minima, a rule-driven error correction to resolve syntax and mapping failures, and a  
  self-learning paradigm that correlates past design decisions with EDA-generated PPA reports to accurately predict future design quality and reduce turnaround time.
  
  
  
  
  \item We comprehensively evaluate MACO against a single-agent LLM design flow, human-expert designs, and ASAP~\cite{tan2022asap} (a state-of-the-art automatic CGRA synthesis framework) across three diverse domains. Under the identical design targets, MACO reduces power consumption by 25.9\%, improves performance by 20.0\%, and accelerates the search process by 5$\times$. Furthermore, we validate MACO-generated designs through a complete ASIC design flow.
  
\end{itemize}

\section{Background and Related Work}

In this section, we discuss the background of CGRAs and the application of LLMs for HW/SW design.

\subsection{Coarse-Grained Reconfigurable
Arrays (CGRAs)}
CGRAs are a class of programmable hardware accelerators that strike a balance between flexibility and efficiency, making them especially well suited for domains such as signal processing, linear algebra, and machine learning. There exist multiple CGRA variants that differ in when and how resources are reconfigured (e.g., spatial, temporal, and hybrid forms)\cite{liu2019survey}. MACO focuses on spatial–temporal CGRAs, where each tile can be reconfigured on a per-cycle basis while computation proceeds in parallel across the array.

\begin{figure}[t]
    \centering
    \includegraphics[width=0.9\linewidth]{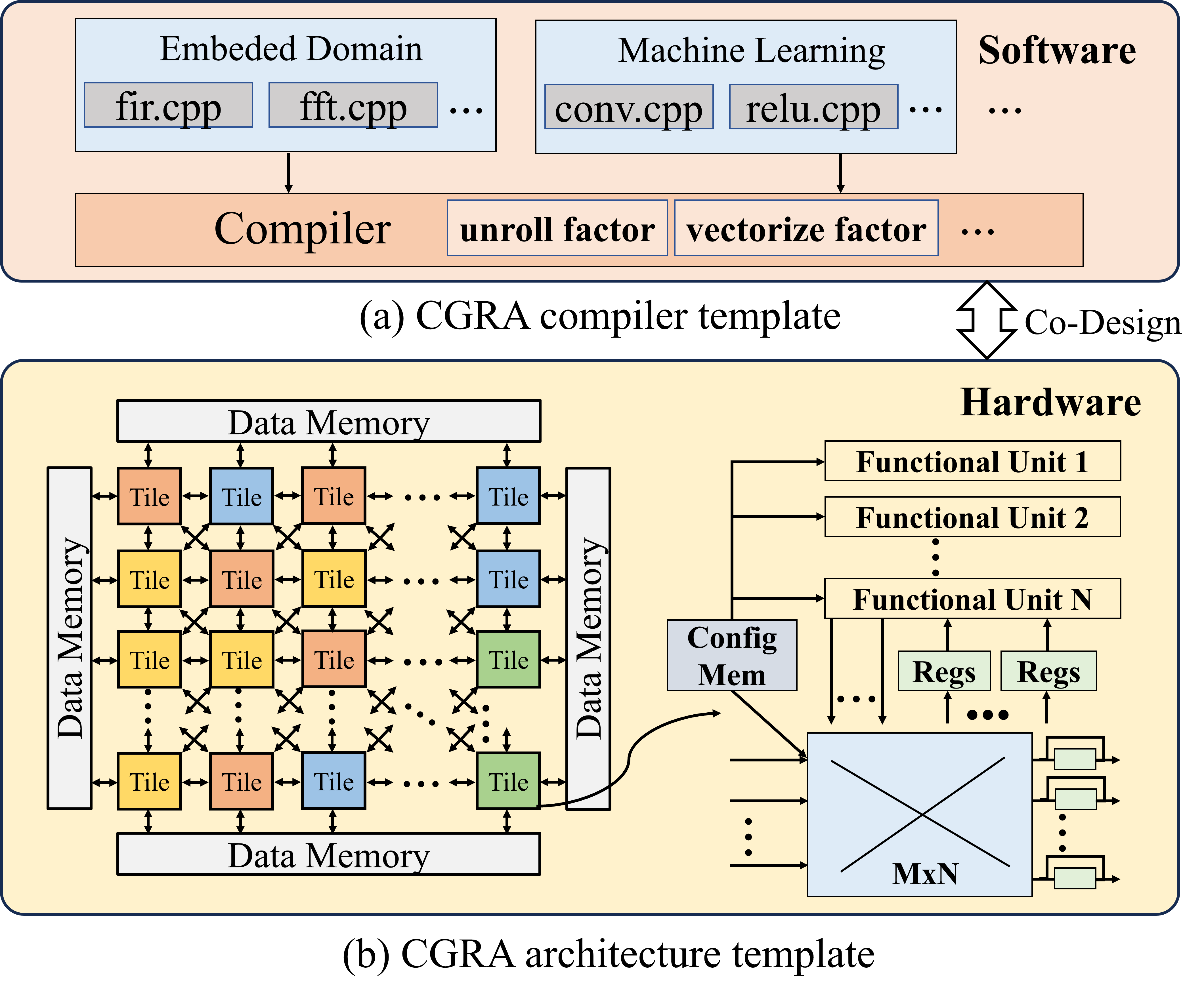}
    \vspace{-10pt}
    \caption{CGRA Compiler and Architecture Design Space.}
    \vspace{-20pt}\label{fig:wide_background}
\end{figure}

The typical CGRA design space, as shown in Fig.~\ref{fig:wide_background}, consists of two main components: compiler (software) design and architecture (hardware) design. A CGRA fabric is organized as a grid of 2D tiles, each integrating functional units, local memory, and interconnects. The compilation software includes choices such as unrolling and vectorization factors, which affect kernel mapping efficiency. Therefore, effective CGRA design requires a joint consideration of both hardware and software parameters.

Due to the inherent complexity of CGRA design, the design process must consider the interdependence among design parameters, performance requirements, and other related factors, which makes identifying an optimal CGRA design both time- and effort-intensive~\cite{weng2019daegen, weng2020dsagen}. By leveraging LLMs, MACO provides a new avenue for addressing these challenges.

\subsection{LLMs in HW/SW Design}
LLMs, such as GPT-5, Qwen, and DeepSeek, have demonstrated remarkable capabilities in processing and generating structured technical content, motivating their application in hardware and software design. They have been employed in 
CPU microarchitecture design~\cite{wang2024chatcpu,cheng2024automated,chang2025large}, the generation of TPU/AI chips~\cite{fu2023gpt4aigchip,vungarala2025tpu,nazzal2025fedchip}, optimization of hardware design flows~\cite{li2025lemoe,zhao2025hardware}, and the construction of hardware-specific datasets for Verilog generation and model fine-tuning~\cite{chang2024data}. Beyond architecture-level work, LLMs also show strong potential for RTL synthesis~\cite{ho2025verilogcoder, thakur2024verigen, liu2023chipnemo}; subsequent efforts improve LLM-based generation via feedback optimization and agent-based methods~\cite{thakur2023autochip, zhao2025mage}, as well as reinforcement learning and multimodal modeling~\cite{wang2025large, chang2024natural, cui2024origen}, with extensions to circuit-level design and benchmarking~\cite{qin2025multi, lu2024rtllm}. However, much of this literature either focuses solely on architectural design without HW/SW co-design or starts directly from RTL, both of which fail to fully consider the interplay between hardware and software during the design process. In addition, some approaches incur high training costs and lack full automation, limiting their ability to achieve optimal CGRA designs.

\textbf{Our Motivation.} 
To address these challenges, we propose a novel LLM-based multi-agent CGRA design flow. Rather than relying on costly large-scale model fine-tuning, MACO employs a holistic HW/SW co-design strategy that iteratively optimizes designs based on direct feedback to maximize overall system performance. Simultaneously, it deploys specialized, autonomous agents to manage distinct tasks—ranging from initial design and verification to error correction and optimization. This multi-agent orchestration streamlines the end-to-end CGRA design process and fundamentally resolves the bottlenecks of previous single-agent approaches. Table~\ref{chip_design_comparsion} compares MACO's capabilities against existing LLM-based chip design methodologies.


\begin{table}[ht]
\centering
\label{tab:comparison}
\footnotesize
\setlength{\tabcolsep}{2pt} 
\renewcommand{\arraystretch}{1.2} 
\begin{tabular}{lcccc}
\toprule
Method & \makecell{HW/SW \\ Co-Design} & \makecell{Iterative \\ Optimization} & {Training-Free} & \makecell Multi-Agent \\
\midrule
ChatCPU~\cite{wang2024chatcpu} & $\times$ & $\times$ & $\checkmark$ &$\checkmark$ \\
FedChip~\cite{nazzal2025fedchip} & $\times$ & $\times$ & $\times$ & $\times$ \\
GPT4AI Chip~\cite{fu2023gpt4aigchip} & $\times$ & $\times$ & $\times$ & $\times$ \\
LPCM~\cite{chang2025large} & $\checkmark$ & $\times$ & $\checkmark$ & $\times$ \\
TPU-gen~\cite{vungarala2025tpu} & $\times$ & $\times$ & $\checkmark$ & $\times$ \\
\textbf{MACO~(Ours)} & $\checkmark$ & $\checkmark$ & $\checkmark$ & $\checkmark$ \\
\bottomrule
\end{tabular}
\vspace{5pt}
\caption{Comparison of MACO with other works.}
\label{chip_design_comparsion}
\vspace{0pt}
\end{table}

\begin{figure*}[t]  
    \centering
    \includegraphics[width=1.0\linewidth]{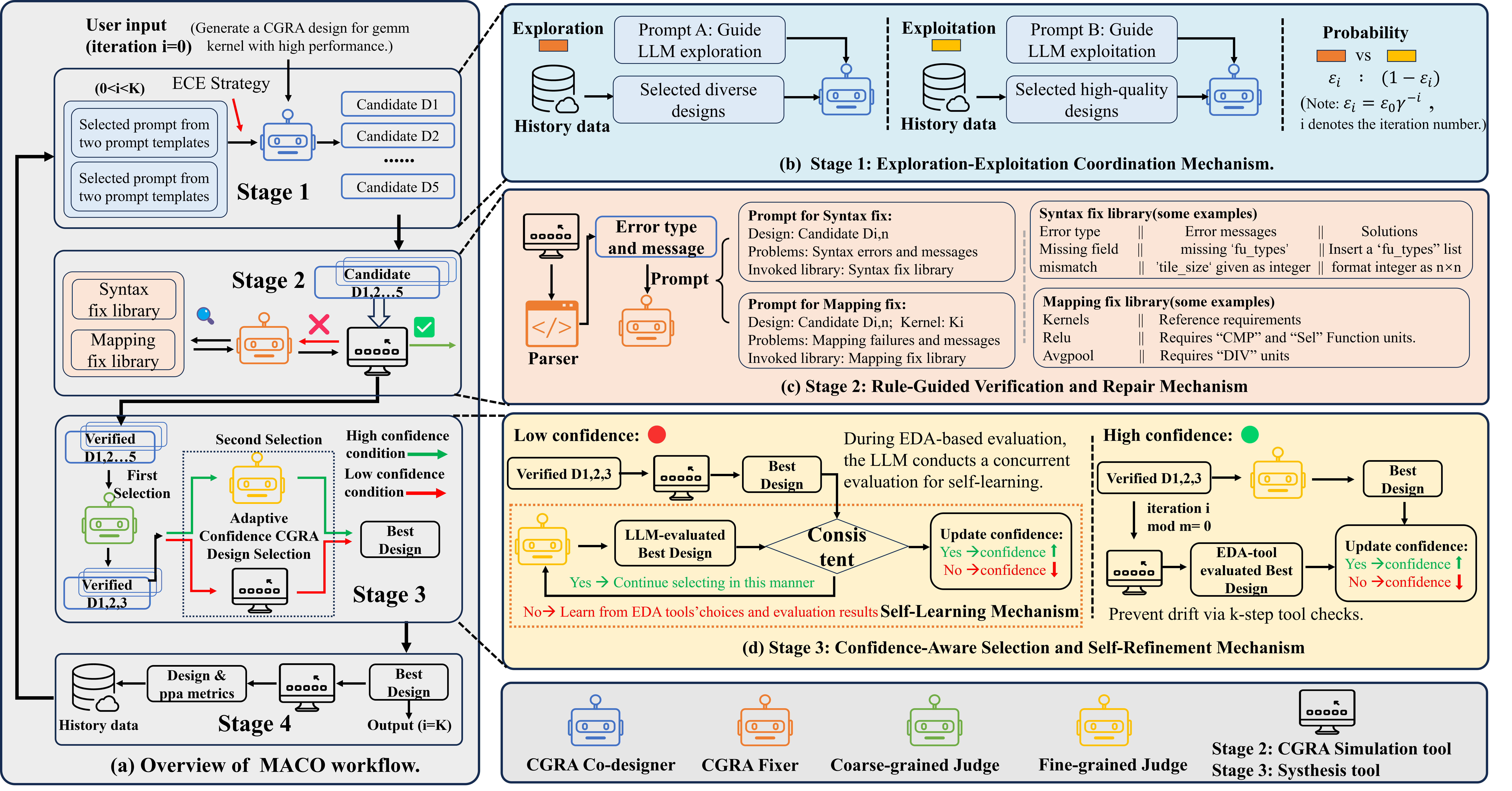}
    \vspace{-15pt}
    \caption{\centering An Overview of the MACO Framework and Details for Stages 1–3.}
    \vspace{-15pt}\label{fig:maco_workflow}
\end{figure*}

\section{MACO Methodology}
This section introduces the MACO, a novel LLM-based multi-agent system tailored for CGRA HW/SW co-design.

\subsection{Overview of the MACO Framework}
\label{sec:overview_maco}
As shown in Fig.~\ref{fig:maco_workflow}, MACO comprises four stages that perform HW/SW co-design, verification and correction, design selection, and evaluation. Each stage is managed by a dedicated agent that executes its specific task while collaboratively completing the overall CGRA design.

\begin{itemize}[leftmargin=1em]
    \item \textbf{Stage 1: CGRA Design} — A domain-aware \textbf{CGRA Co-designer} with HW/SW co-design principles and CGRA domain knowledge generates high-quality candidate designs encompassing both hardware and software parameters (e.g., functional unit allocation and loop unrolling). To efficiently navigate this joint space, the agent leverages tailored prompts and historical data via an \textbf{Exploration-Exploitation Coordination Mechanism} (detailed in Sec.~\ref{sec3.2}) to guide the LLM's generation.

    \item \textbf{Stage 2: Validation \& Correction} — This stage validates CGRA candidate designs using simulation/verification tools~\cite{tan2020opencgra}. Designs that pass proceed directly to the next stage; designs that fail are repaired through k-th rounds by \textbf{CGRA  Fixer} with \textbf{Rule-Driven Repair Mechanism}. Moreover, we construct a \textit{Syntax Fix Library} and a \textit{Mapping Fix Library} that catalog common errors and remedies to support LLM-driven repair; see Sec.~\ref{sec:rule_knowledge_repair} for details.
    \item \textbf{Stage 3: Multi-Judge Selection} — Verified designs are screened via a two-level hierarchical process. At the first level, a \textbf{coarse-grained judge} leverages the LLM’s reasoning and embedded optimization knowledge to quickly identify candidates with high performance potential. At the second level, a \textbf{fine-grained judge} collaborates with CGRA design and synthesis tools to select the best CGRA design under an adaptive-confidence scheme: in early iterations, synthesis tools provide accurate power/area analysis while the open-source CGRA tool~\cite{tan2020opencgra} estimates performance, enabling selection of the best CGRA design. During this process, the LLM, although not making the final decision, still produces its own choice and learns by comparing it with the EDA-tool-based results. As iterations progress, the \textbf{fine-grained judge} learns from experimental data and evolves into an autonomous evaluator capable of replacing the tools to make the final design selection. This procedure follows an approach called \textbf{Confidence-Adaptive Selection and Self-learning Mechanism}, which is discussed in Sec.~\ref{sec3.4}.
    \item \textbf{Stage 4: Performance Evaluation \& Feedback} — Final designs that were not fully evaluated in Stage~3 undergo comprehensive performance and power assessment using OpenCGRA~\cite{tan2020opencgra, tan2021opencgra} and synthesis tools. The results are stored as historical data and fed back to the Stage 1 to further improve the co-designer’s generation quality. Finally, MACO outputs the final design results.
\end{itemize}

Together, these four stages form a closed-loop iterative design process that reduces manual intervention and significantly accelerates CGRA design and optimization.

\begin{figure}[t]  
    \centering
    \includegraphics[width=1.0\linewidth]{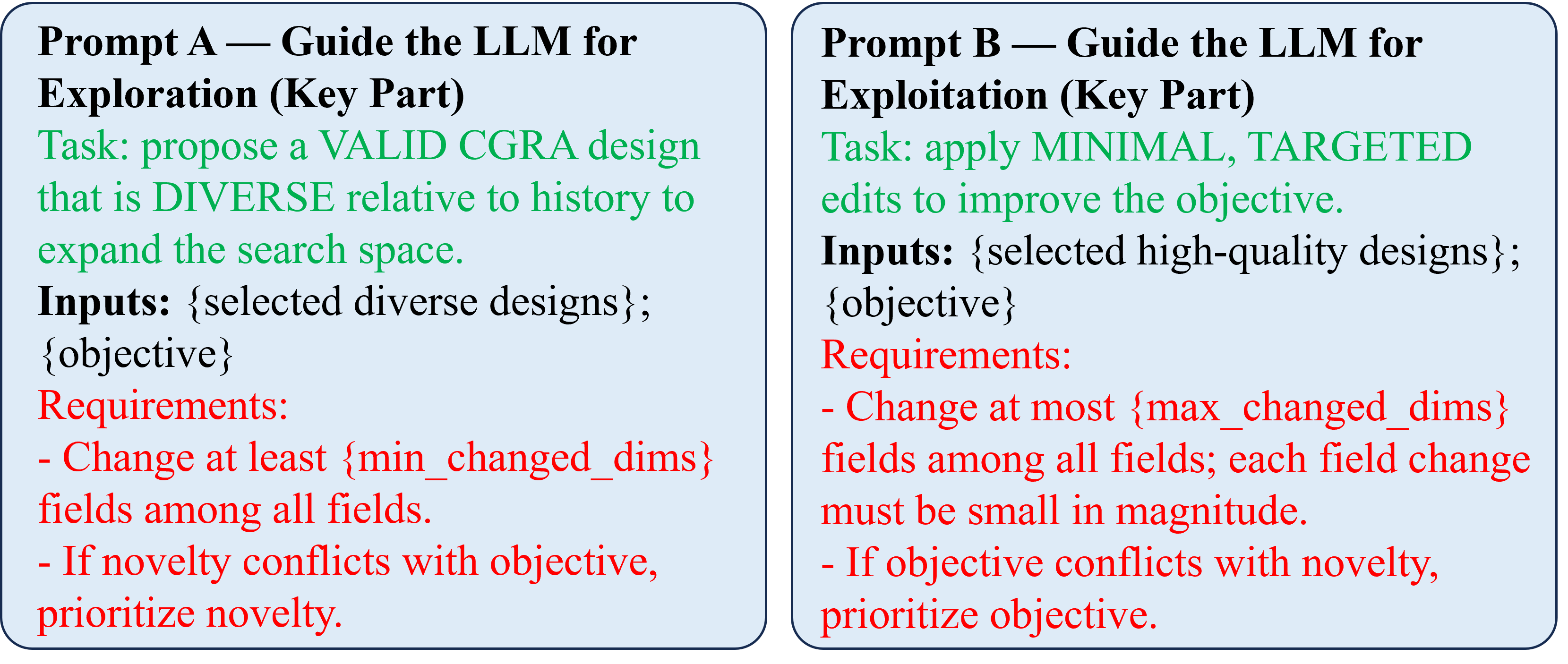}
    \vspace{-15pt}
    \caption{Prompts for balancing  exploration and exploitation.}
    \label{fig:promptAB}
    \vspace{-10pt}
\end{figure}

\subsection{Stage-1 Exploration-Exploitation Coordination Mechanism}
\label{sec3.2}

During our experiments, we observed that LLMs tend to increasingly rely on historical designs and tool reports when exploring the CGRA design space. Without explicit guidance on where to explore next, this history-driven refinement can gradually bias the search toward previously seen design patterns, thereby limiting exploration efficiency and hampering the discovery of better designs. To address this issue, we introduce a control knob that explicitly regulates the agent's exploration--exploitation behavior, rather than allowing the search to drift into a self-generated-data dependency loop. Specifically, we adopt an exponentially decaying $\varepsilon$-greedy strategy, where $\varepsilon$ denotes the exploration rate and gradually decreases over iterations to balance exploration and exploitation, as illustrated in Figs.~\ref{fig:maco_workflow}(a) and (c).

As shown in Eqs.~\ref{eq:1}--\ref{eq:2}, an exponentially decaying probability parameter governs whether the LLM explores or exploits (see Fig.~\ref{fig:maco_workflow}(c)). As iterations progress, $\varepsilon_{i,n}$ decreases monotonically, shifting the policy from high-probability exploration to high-probability exploitation. This aligns with traditional heuristic DSE algorithms~\cite{jing2013application}, in which a probabilistic parameter $\varepsilon$ is used to ensure adequate design-space exploration.


\begin{equation}
\vspace{-15pt}
\varepsilon_{i,n}=\varepsilon_0\,\gamma^{\,i-1}, \ for \  n=1,\dots,N_{\mathrm{cand}},\  in \ i=1,\dots, K.
\label{eq:1}
\vspace{-8pt}
\end{equation}

\begin{equation}
D_{i,n} =
\begin{cases}
\mathrm{LLM\_Exploration}\!\left(P_A, S_\text{highVar}\right), & \text{if } u_{i,n} < \varepsilon_{i,n},\\
\mathrm{LLM\_Exploitation}\!\left(P_B, S_\text{highPerf}\right), & \text{if } u_{i,n} \ge \varepsilon_{i,n}.
\end{cases}
\label{eq:2}
\end{equation}

Here, $i=1,\dots,K$ indexes the iteration and $n=1,\dots,N_{\mathrm{cand}}$ indexes the candidate within each iteration. 
$\varepsilon_0$ denotes the initial exploration rate and $\gamma\in(0,1)$ is the decay factor; $\varepsilon_{i,n}$ decays geometrically. The random variable $u_{i,n}\sim\mathcal{U}(0,1)$ determines whether a candidate is generated via exploration or exploitation. 

\begin{figure}[t]
    \centering
    \includegraphics[width=1.0\linewidth]{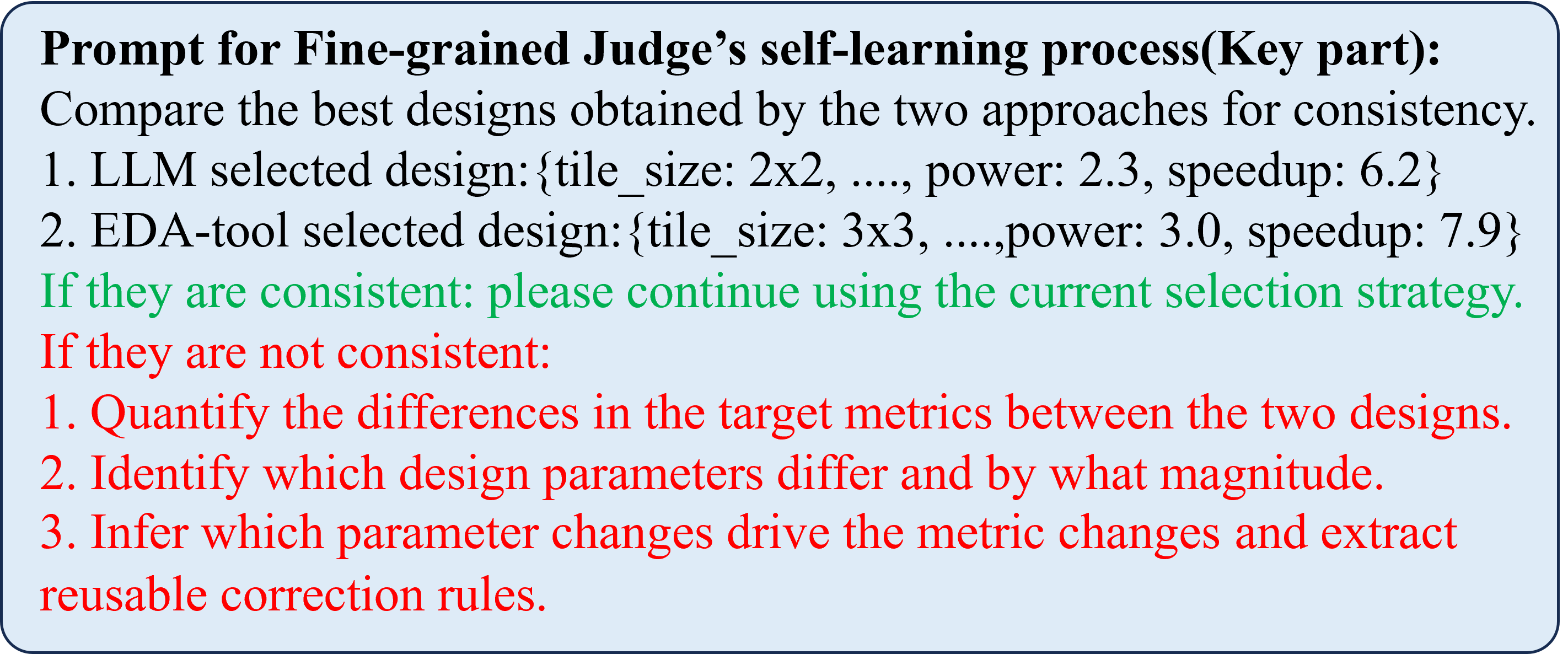}
    \vspace{-15pt}
    \caption{Prompt design for fine-grained judge's self-learning.}
    \label{fig:self_learning}
    \vspace{-10pt}
\end{figure}




We pair the schedule with two prompt templates ($P_A$ and $P_B$) -- one for exploration and the other for exploitation, which is shown in Fig.~\ref{fig:promptAB}, and
seed them (respectively) with two curated historical subsets: a \emph{high-variance} set (diverse designs) and a
\emph{top-$k$ best-performing} set (high-quality designs) based on the target objective, to generate CGRA candidates. As shown by the ablation study in Fig.~\ref{fig:ablation-decaying} (a), this strategy effectively improves the final design quality of MACO.

\subsection{Stage-2 Rule-Guided Repair Mechanism}
\label{sec:rule_knowledge_repair}
In Stage~2, as shown in Figs.~\ref{fig:maco_workflow}(a) and \ref{fig:maco_workflow}(d), the open-source toolchain \textit{OpenCGRA}~\cite{tan2020opencgra,tan2021opencgra} is used to validate CGRA designs generated by MACO. When syntax or mapping errors are detected, the parser processes the tool-generated reports to determine the error type and extract salient error information. Based on the identified type, it selects an appropriate prompt to feed the agent, enabling the LLM to quickly localize the fault type and query the right library to resolve it.

In our experiments, observed generation errors generally fall into two categories: \emph{syntax errors} and \emph{mapping failures}. 
To assist the LLM in repair, \textbf{CGRA Fixer} introduces two specialized libraries. The \textit{Syntax Fix Library} provides curated remedies for easily localized issues, such as missing fields or type mismatches. 
Conversely, the \textit{Mapping Fix Library} addresses complex, application-specific bottlenecks, like missing functional units or insufficient memory banks, by supplying the LLM with typical CGRA architectural requirements (Fig.~\ref{fig:maco_workflow}(d)). 
As shown in Table~\ref{tab:pass1_avg_models}, integrating \textbf{CGRA Fixer} across three LLMs improves the average valid-design rate by 25.5\%, significantly reducing hallucinations and enhancing overall robustness.

\subsection{Stage-3 Confidence-Adaptive Selection and Self-Learning}
\label{sec3.4}
In Stage 3, as shown in Figs.~\ref{fig:maco_workflow}(a) and \ref{fig:maco_workflow}(d), when selecting the optimal design from the top-$K$ candidate designs, we we evaluate two primary approaches: rigorous EDA tool evaluation and rapid inference via the Fine-Grained Judge. To effectively navigate the trade-off between the time-consuming nature of tool-based evaluation and the inherent reasoning limitations of LLMs, we propose a confidence-adaptive selection and self-learning mechanism designed to significantly reduce overall turnaround time without compromising design quality.

At the beginning of the iterative process, a baseline confidence score and a predefined trust threshold are initialized for the fine-grained judge.
During each iteration, if the LLM's confidence remains below this threshold, EDA tools (e.g., simulation and synthesis) are 
invoked to evaluate the top-$k$ candidate designs, and the final selection is dictated by these tool-derived ground-truth results. 
Concurrently, the LLM generates its own independent prediction and updates its internal judgment model by comparing its selection against the authoritative EDA reports.
As the iterations proceed, the LLM's confidence score is dynamically updated based on the historical consistency between its predictions and the tool-based evaluations.
Once this confidence score exceeds the trust threshold, MACO increasingly relies on the LLM's reasoning for design selection, thereby significantly mitigating the computational cost of repeated EDA tool invocations.
To guarantee continued robustness and prevent model drift, EDA tool-based evaluations are periodically reintroduced every $m$ iterations for recalibration. 
As illustrated in Fig.~\ref{fig:self_learning}, this iterative process enables the fine-grained judge to progressively refine its design-selection capability utilizing both direct EDA tool feedback and its own historical decisions.
Fig.~\ref{fig:ablation-adaptive-combined}(b) and Fig.~\ref{fig:6}(b) 
show that this adaptive mechanism effectively reduces the overall PPA evaluation time compared to a strictly tool-only selection baseline, while maintaining design quality.

\definecolor{ModelA}{RGB}{246,183,198}   
\definecolor{ModelB}{RGB}{162,218,222}    
\definecolor{ModelC}{RGB}{254,175,138}    
\input{5-sec-evaluation}


\section{Conclusion}
This paper presents MACO, the first multi-agent LLM framework for automated CGRA hardware/software co-design. By orchestrating specialized agents equipped with a dynamic exploration strategy and EDA-guided self-learning, MACO effectively navigates the vast joint design space. Compared to state-of-the-art baselines, MACO reduces power consumption by 25.9\%, improves performance by 20.0\%, and accelerates the search process by 5$\times$. Future work will extend MACO to support more complex architectures and integrate real-silicon deployment feedback.

\bibliographystyle{unsrt}
\scriptsize
\bibliography{ref}
\end{document}